# Energy Conservation Equations of Motion

N. A. Vinokurov

Budker Institute of Nuclear Physics, Siberian Branch of the Russian Academy of Sciences, prosp. Akademika Lavrent'eva 11, 630090 Novosibirsk, Russian Federation,
e-mail: vinokurov@inp.nsk.su;
Korea Atomic Energy Research Institute, 1045 Daedeok-Daero, Yuseong-gu, Daejeon 305-353, Republic of Korea

**Abstract**
A conventional derivation of motion equations in mechanics and field equations in field theory is based on the principle of least action with a proper Lagrangian. With a time-independent Lagrangian, a function of coordinates and velocities that is called energy is constant. This paper presents an alternative approach, namely derivation of a general form of equations of motion that keep the system energy, expressed as a function of generalized coordinates and corresponding velocities, constant. These are Lagrange's equations with addition of gyroscopic forces. The important fact, that the energy is defined as the function on the tangent bundle of configuration manifold, is used explicitly for the derivation. The Lagrangian is derived from a known energy function. A development of generalized Hamilton's and Lagrange's equations without the use of variational principles is proposed. The use of new technique is applied to derivation of some equations.

## 1. Introduction

For a wide class of mechanical systems and fields, equations of motion can be derived from variational principles (see, e. g., [1 – 5]). On the one hand, this fact may reflect some specific properties of related differential equations and can be considered as a more compact writing of these equations. On the other hand, these specific properties and variation principles are important since they form the background of quantum mechanics. In these approaches, conservation of energy is a consequence of no explicit dependence of the Lagrangian on time. This paper suggests an alternative approach. Based on conservation of energy, we derive equations of motion that are a generalization of Lagrange's equations. The important fact, that the energy is defined as the function on the tangent bundle (of coordinates and velocities), is used explicitly for the derivation.

## 2. Generalized Lagrange's equations

Let $E(\mathbf{q},\mathbf{v}) = const$ be the energy of the system, which depends on the generalized coordinates $q^i$, $i = 1,\ldots N$ and generalized velocities $v^i = dq^i/dt$, the bold letters denoting a set of variables: $\mathbf{q} = (q^1, \ldots, q^N)$. The state of the system is described with a point with the coordinates $(\mathbf{q}, \mathbf{v})$ on the tangent bundle $\mathbf{TQ}$ of $N$-dimension manifold $\mathbf{Q} = \{\mathbf{q}\}$ [5]. Conservation of energy along the motion trajectory can be written as

$$0 = \frac{dE(\mathbf{q},\mathbf{v})}{dt} = \frac{\partial E}{\partial q^i} v^i + \frac{\partial E}{\partial v^i} \dot{v}^i \qquad (1)$$

Now we will try to find equations of motion that conserve the energy. Eq. (1) needs to be true for any velocity $\mathbf{v}$. Then Eq. (1) is identity. One scalar identity $v^i f_i(\mathbf{q},\mathbf{v}) = 0$ in the tangent bundle **TQ** is equivalent to $N$ identities $f_i(\mathbf{q},\mathbf{v}) = C_{ij}(\mathbf{q},\mathbf{v})v^j$ for any antisymmetric tensor $C_{ij}$. To use it one can expand the energy in a Taylor's series,

$$E(\mathbf{q},\mathbf{v}) = \sum_{n=0}^{\infty} A^{(n)}_{ik...l}(\mathbf{q}) v^i v^k ... v^l. \tag{2}$$

Substitution of Eq. (2) to Eq. (1) gives

$$0 = v^i \sum_{n=0}^{\infty} \frac{\partial A^{(n)}_{jk...l}(\mathbf{q})}{\partial q^i} v^j v^k ... v^l + \dot{v}^i \frac{\partial}{\partial v^i} \sum_{n=0}^{\infty} A^{(n)}_{jk...l}(\mathbf{q}) v^j v^k ... v^l$$

$$= v^i \frac{\partial A^{(0)}(\mathbf{q})}{\partial q^i} + v^i v^j \frac{\partial A^{(1)}_j(\mathbf{q})}{\partial q^i} + \dot{v}^i A^{(1)}_i(\mathbf{q}) + v^i \sum_{n=2}^{\infty} \frac{\partial A^{(n)}_{jk...l}(\mathbf{q})}{\partial q^i} v^j v^k ... v^l + \dot{v}^i \frac{\partial}{\partial v^i} \sum_{n=2}^{\infty} A^{(n)}_{jk...l}(\mathbf{q}) v^j v^k ... v^l. \tag{3}$$

$$= \dot{v}^i A^{(1)}_i(\mathbf{q})$$

$$+ v^i \left[ \frac{\partial A^{(0)}(\mathbf{q})}{\partial q^i} + v^j \frac{\partial A^{(1)}_j(\mathbf{q})}{\partial q^i} + \sum_{n=2}^{\infty} \frac{\partial A^{(n)}_{jk...l}(\mathbf{q})}{\partial q^i} v^j v^k ... v^l + \dot{v}^m \frac{\partial^2}{\partial v^i \partial v^m} \sum_{n=2}^{\infty} \frac{1}{n-1} A^{(n)}_{jk...l}(\mathbf{q}) v^j v^k ... v^l \right]$$

Most of the physically interesting cases have no linear terms in energy expansion Eq. (2). Therefore we will limit our further consideration with the condition $A^{(1)}_i(\mathbf{q}) = 0$. To simplify Eq. (3), one can define the following function:

$$L_1 = \sum_{n \neq 1} \frac{1}{n-1} A^{(n)}_{ij...k} v^i v^j ... v^k = -A^{(0)} + \sum_{n=2}^{\infty} \frac{1}{n-1} A^{(n)}_{ij...k} v^i v^j ... v^k$$

$$= -E(\mathbf{q},0) + \int_1^{\infty} \left[ E\left(\mathbf{q}, \frac{\mathbf{v}}{x}\right) - E(\mathbf{q},0) \right] dx \tag{4}$$

It is easy to check that it satisfies the partial differential equation:

$$\frac{\partial L(\mathbf{q},\mathbf{v})}{\partial v^i} v^i - L(\mathbf{q},\mathbf{v}) = E(\mathbf{q},\mathbf{v}). \tag{5}$$

In textbooks on mechanics (see, e. g., [5]), Eq. (5) is used for definition of the energy $E$ through a given Lagrangian $L$. Here, we have to find $L$ via solving partial differential equation (5). A complete solution of Eq. (5) can be written as a sum of a particular solution $L_1$ and a complete solution of the corresponding homogeneous equation:

$$L = L_1(\mathbf{q},\mathbf{v}) + a_i(\mathbf{q}) v^i. \tag{6}$$

The second term of Eq. (6) includes $N$ arbitrary constants $a_i$ and describes gyroscopic-like forces, which do not contribute to the energy $E(\mathbf{q},\mathbf{v})$. It is worth noting that any function of energy $F[E(\mathbf{q},\mathbf{v})]$ can be used in Eq. (5) instead of energy $E(\mathbf{q},\mathbf{v})$. Other Lagrangians can then be found with Eq. (4) and Eq. (6). A simple example is $E(\mathbf{q},\mathbf{v}) = A^{(0)} + A^{(2)}_{ik} v^i v^k = U(\mathbf{q}) + T(\mathbf{q},\mathbf{v})$ and $F(E) = E^2$. In this case, Eq. (4) gives $L_1 = T^2/3 + 2UT - U^2$.

With the use of Lagrange's function $L_1$, Eq. (3) takes the following form:

$$0 = v^i \left[ v^m \frac{\partial^2}{\partial q^m \partial v^i} \sum_{n \neq 1}^{\infty} \frac{A^{(n)}_{jk...l}(\mathbf{q})}{n-1} v^j v^k ... v^l - \frac{\partial}{\partial q^i} \sum_{n \neq 1}^{\infty} \frac{A^{(n)}_{jk...l}(\mathbf{q})}{n-1} v^j v^k ... v^l \right.$$

$$\left. + \dot{v}^m \frac{\partial^2}{\partial v^i \partial v^m} \sum_{n \neq 1}^{\infty} \frac{1}{n-1} A^{(n)}_{pj...l}(\mathbf{q}) v^p v^j ... v^l \right]$$

$$= v^i \left[ -\frac{\partial}{\partial q^i} L_1(\mathbf{q},\mathbf{v}) + v^m \frac{\partial^2}{\partial q^m \partial v^i} L_1(\mathbf{q},\mathbf{v}) + \dot{v}^m \frac{\partial^2}{\partial v^i \partial v^m} L_1(\mathbf{q},\mathbf{v}) \right] \tag{7}$$

$$= v^i \left[ \frac{d}{dt} \frac{\partial L_1(\mathbf{q},\mathbf{v})}{\partial v^i} - \frac{\partial}{\partial q^i} L_1(\mathbf{q},\mathbf{v}) \right]$$

As **v** is arbitrary, Eq. (7) is equivalent to generalized Lagrange's equations

$$\frac{d}{dt} \frac{\partial L_1}{\partial v^i} - \frac{\partial L_1}{\partial q^i} = G_{ij}(\mathbf{q},\mathbf{v},\dot{\mathbf{v}},t) v^j, \tag{8}$$

where $G_{ij} = -G_{ji}$ is an antisymmetric tensor. It can also be written as

$$\frac{d}{dt}\frac{\partial L}{\partial v^i} - \frac{\partial L}{\partial q^i} = \left[ G_{ij}(\mathbf{q},\mathbf{v},\dot{\mathbf{v}},t) + \frac{\partial a_i(\mathbf{q})}{\partial q^j} - \frac{\partial a_j(\mathbf{q})}{\partial q^i} \right] v^j = C_{ij}(\mathbf{q},\mathbf{v},\dot{\mathbf{v}},t) v^j, \qquad (9)$$

where $C_{ij}$ is an antisymmetric tensor.

Generalized Lagrange's equations (8) present equations of motion that satisfy the energy conservation law with the given energy function $E(\mathbf{q},\mathbf{v})$. The right-hand side of Eq. (8) represents generalized gyroscopic forces, which do not change the system energy. In general case differential form $\mathbf{G}$ is not exact. Then the generalized gyroscopic forces can not be included in Lagrangian with some functions $a_i$. Therefore one can conclude that energy conservation allows existence of forces, which can not be described by standard Lagrange's equations. It is worth noting that generalized Lagrange's equations were obtained without use of and, in general case, do not meet any variational principle. Eq. (8) was mentioned first without derivation in [6] to underline, that Lagrange's equations are not the general motion equations to meet the energy conservation.

## 3. Generalized Hamilton's equations

The standard Legendre transform of $L$,

$$L(\mathbf{q},\mathbf{v}) + H(\mathbf{q},\mathbf{p}) = p_i v^i, \qquad (10)$$

transforms Eq. (9) to generalized Hamilton's equations

$$\frac{d}{dt}p_i = -\frac{\partial H}{\partial q^i} + C_{ij}\left(\mathbf{q}, \frac{\partial H}{\partial \mathbf{p}}, \frac{\partial H}{\partial p_k}\frac{\partial^2 H}{\partial \mathbf{p}\partial q^k} - \frac{\partial H}{\partial q^k}\frac{\partial^2 H}{\partial \mathbf{p}\partial p_k}, t\right)\frac{\partial H}{\partial p_j} = \{p_i, H\}$$
$$\frac{d}{dt}q^i = \frac{\partial H}{\partial p_i} = \{q^i, H\} \qquad (11)$$

with generalized momenta

$$p_i = \frac{\partial L}{\partial v^i} = \frac{\partial L_1}{\partial v^i} + a_i = \int_1^\infty \frac{\partial}{\partial v^i} E\left(\mathbf{q}, \frac{\mathbf{v}}{x}\right) dx + a_i, \qquad (12)$$

Hamiltonian $H = E[\mathbf{q},\mathbf{v}(\mathbf{p},\mathbf{q})]$ and the bracket

$$\{f,g\} = \frac{\partial f}{\partial q^i}\frac{\partial g}{\partial p_i} - \frac{\partial g}{\partial q^i}\frac{\partial f}{\partial p_i} + C_{ij}\frac{\partial f}{\partial p_i}\frac{\partial g}{\partial p_j}. \qquad (13)$$

As it is clear from Eq. (11, 13), the $2N \times 2N$ matrix in the bracket is

$$\mathbf{J} = \begin{pmatrix} \|C_{ij}\| & -\mathbf{E} \\ \mathbf{E} & 0 \end{pmatrix}, \qquad (14)$$

where $\mathbf{E}$ is a $N \times N$ unity matrix. Eq. (13) is Poisson's bracket if it satisfies the Jacobi identity [5, 7]

$$\sum_{\delta=1}^{2N}\left( J^{\alpha\delta}\frac{\partial J^{\beta\gamma}}{\partial x^\delta} + J^{\gamma\delta}\frac{\partial J^{\alpha\beta}}{\partial x^\delta} + J^{\beta\delta}\frac{\partial J^{\gamma\alpha}}{\partial x^\delta} \right) = 0, \qquad (15)$$

where the Greek indices vary from 1 to $2N$ and $x_i = p_i$; $x_{N+i} = q^i$. These conditions give

$$\frac{\partial C_{ij}}{\partial p_k} = 0 \qquad (16)$$

and

$$\frac{\partial C_{jk}}{\partial q^i} + \frac{\partial C_{ij}}{\partial q^k} + \frac{\partial C_{ki}}{\partial q^j} = 0. \qquad (17)$$

The last equation means that differential form $\mathbf{C}$ is closed on manifold $\mathbf{Q}$. In three-dimensional space using Levi-Civita symbol $e_{ijk}$ one can express form $\mathbf{C}$ through vector:

$$C_{ij} = e_{ijk}B^k. \qquad (18)$$

Then Eq. (17) is equivalent to

$$\frac{\partial B^k}{\partial q^k} = 0. \qquad (19)$$

By Helmholtz theorem, Eq. (19) allows to express $B^i = e^{ijk} \partial A_k/\partial q^j$ through "vector potential" **A**. Then

$$C_{ij} = \frac{\partial A_j}{\partial q^i} - \frac{\partial A_i}{\partial q^j}, \qquad (20)$$

i. e., form **C** is exact and can be reduced to zero by adding term **vA** to Lagrangian.

In general case, there is a Poisson's structure if generalized Lagrange's equation (9) can be written as

$$\frac{d}{dt}\frac{\partial L}{\partial v^i} - \frac{\partial L}{\partial q^i} = \left[C_{ij}\left(\frac{\partial L}{\partial \mathbf{v}} - \mathbf{a}\right) + \frac{\partial a_i(\mathbf{q})}{\partial q^j} - \frac{\partial a_j(\mathbf{q})}{\partial q^i}\right]v^j, \qquad (21)$$

and the conditions of Eq. (17) are satisfied.

Thus, postulating conservation of a known function $E(\mathbf{q},\mathbf{v})$ of generalized coordinates and velocities, one can derive (with some ambiguity caused by the possibility of gyroscopic-type forces) generalized Lagrange's and then Hamilton's equations. As the energy conservation is one of the most basic lows of physics, many motion equations can be derived using generalized Lagrange equations Eq. (8). Some of them are discussed below.

## 4. Hamilton's and Schrödinger's equations

There is an interesting example of energy $E(\mathbf{q}) = U(\mathbf{q})$ that does not depend on the velocities. Then the general solution of Eq. (5) is $L = -U(\mathbf{q}) + a_i(\mathbf{q})v^i$ and generalized Lagrange's equations (8) give

$$\frac{\partial U(\mathbf{q})}{\partial q^i} = G_{ij}(\mathbf{q},\mathbf{v},\dot{\mathbf{v}},t)v^j. \qquad (22)$$

The case when $G_{ij}$ does not depend on velocities has to be discussed separately, as the initial conditions of a system of the first-order differential equations are given as a point **q** in manifold **Q**. Then, according to Eq. (1), velocity vector **v** is tangent to the surface of constant energy, and therefore is orthogonal to the energy gradient:

$$\frac{dq^i}{dt} = J^{ij}(\mathbf{q},\mathbf{v},t)\frac{\partial U(\mathbf{q})}{\partial q^j}. \qquad (23)$$

where $J^{ij}$ is arbitrary antisymmetric tensor. Eq. (23) are Hamiltonian if $J^{ij}$ depends only on **q** and satisfies Jacobi identity, similar to Eq. (15), but with summation from 1 to $N$. It is worth noting that the dimension $N$ of the configuration space, may be odd here.

In a particular case of a quadratic dependence $U(\mathbf{q}) = h_{ij}q^iq^j$ ($h_{ij} = h_{ji}$) and constant (independent on **q**) $J^{ij}$, Eq. (23) is linear:

$$\frac{dq^i}{dt} = 2J^{ij}h_{jk}q^k. \qquad (24)$$

Let this equation conserve the "length" $\sqrt{g_{ij}q^iq^j} = 1$ of the vector **q** with constant metric tensor **g**:

$$0 = \frac{d}{dt}g_{ij}q^iq^j = 4g_{ij}J^{jk}h_{kl}q^iq^l. \qquad (25)$$

That means that the tensor **gJh** in the right side of Eq. (25) needs to be antisymmetric. Below we will suppose that matrix of **g** is the unity one, i. e., the norm of vector **q** is Euclidian. Then $\mathbf{Jh} = -(\mathbf{Jh})^T = \mathbf{hJ}$.

For even $N = 2M$, the proper orthogonal transformation of the coordinates can reduce **J** to

$$\mathbf{J} = \frac{1}{2}\begin{pmatrix} \mathbf{0} & \mathbf{D} \\ -\mathbf{D} & \mathbf{0} \end{pmatrix} \qquad (26)$$

with any non-singular diagonal square matrix **D**. Representing the symmetric matrix **h** as four blocks, one can rewrite the commutation relation as follows:

$$\begin{pmatrix} \mathbf{0} & \mathbf{D} \\ \mathbf{-D} & \mathbf{0} \end{pmatrix} \begin{pmatrix} \mathbf{H}_1 & \mathbf{H}_2^T \\ \mathbf{H}_2 & \mathbf{H}_3 \end{pmatrix} = \begin{pmatrix} \mathbf{DH}_2 & \mathbf{DH}_3 \\ \mathbf{-DH}_1 & \mathbf{-DH}_2^T \end{pmatrix} = \begin{pmatrix} \mathbf{H}_1 & \mathbf{H}_2^T \\ \mathbf{H}_2 & \mathbf{H}_3 \end{pmatrix} \begin{pmatrix} \mathbf{0} & \mathbf{D} \\ \mathbf{-D} & \mathbf{0} \end{pmatrix} = \begin{pmatrix} \mathbf{-H}_2^T \mathbf{D} & \mathbf{H}_1 \mathbf{D} \\ \mathbf{-H}_3 \mathbf{D} & \mathbf{H}_2 \mathbf{D} \end{pmatrix}. \tag{27}$$

Therefore $\mathbf{H}_1\mathbf{D} = \mathbf{DH}_3$, $\mathbf{DH}_2 = -(\mathbf{DH}_2)^T$ and $\mathbf{H}_2\mathbf{D} = -(\mathbf{H}_2\mathbf{D})^T$. Two last equations give $\mathbf{D}^2\mathbf{H}_2 = \mathbf{H}_2\mathbf{D}^2$. To satisfy this condition for arbitrary $\mathbf{H}_2$, $\mathbf{D}^2$ has to be proportional to the unity matrix. Then $D_{ij} = \pm\kappa\delta_{ij}$. Proper permutation of coordinates $q^j$ and $q^{M+j}$ provides $D_{ij} = \kappa\delta_{ij}$. Then $\mathbf{H}_1 = \mathbf{H}_3$, $\mathbf{H}_2 = -\mathbf{H}_2^T$, and matrix $\mathbf{H} = \mathbf{H}_1 + i\mathbf{H}_2$ is Hermitian. Defining a complex vector $\psi^j = q^j + iq^{M+j}$ for $j = 1, \ldots M$, one can rewrite Eq. (24) as follows:

$$i\frac{d\boldsymbol{\psi}}{dt} = \kappa\mathbf{H}\boldsymbol{\psi}. \tag{28}$$

This is Schrödinger's equation, obtained from the conservation of the norm (total probability) $\boldsymbol{\psi}^\dagger\boldsymbol{\psi} = 1$ and the average energy $U(\mathbf{q}) = h_{ij}q^i q^j = \boldsymbol{\psi}^\dagger \mathbf{H}\boldsymbol{\psi}$. Here $\boldsymbol{\psi} = (\psi^1, \ldots, \psi^M)^T$ is a column and $\boldsymbol{\psi}^\dagger = (\psi^{1*}, \ldots, \psi^{M*})$ is a row of complex conjugate values.

## 5. Application to fields

The above technique can be applied to a field. In this case, the degree of freedom index $i$ is replaced with the observation point coordinate $\mathbf{r}$. Let, for example, $\varphi(\mathbf{r},t)$ be a scalar field in a three-dimensional space and

$$E\left(\varphi, \frac{\partial\varphi}{\partial t}\right) = \frac{1}{2}\int\left[\left(\frac{\partial\varphi}{\partial t}\right)^2 + c^2(\nabla\varphi)^2 + m^2\varphi^2\right]d^3\mathbf{r}. \tag{29}$$

Therefore, the derivatives with respect to $\varphi$ and $\partial\varphi/\partial t$ in the above equations need to be variational. According to Eq. (4),

$$L_1\left(\varphi, \frac{\partial\varphi}{\partial t}\right) = \frac{1}{2}\int\left[\left(\frac{\partial\varphi}{\partial t}\right)^2 - c^2(\nabla\varphi)^2 - m^2\varphi^2\right]d^3\mathbf{r}. \tag{30}$$

Then Eq. (8) gives

$$\frac{\partial^2\varphi}{\partial t^2} - c^2\Delta\varphi + m^2\varphi = \int G(\varphi,\mathbf{r},\mathbf{r}',t)\frac{\partial\varphi(\mathbf{r}',t)}{\partial t}d^3\mathbf{r}' \tag{31}$$

with $G(\varphi,\mathbf{r},\mathbf{r}',t) = -G(\varphi,\mathbf{r}',\mathbf{r},t)$. For local interaction,

$$G(\varphi,\mathbf{r},\mathbf{r}',t) = K\left(\varphi, \frac{\mathbf{r}+\mathbf{r}'}{2}, \mathbf{r}-\mathbf{r}',t\right) =$$
$$f_i\left(\varphi, \frac{\mathbf{r}+\mathbf{r}'}{2}\right)\frac{\partial}{\partial r_i}\delta(\mathbf{r}-\mathbf{r}') + g_{ijk}\left(\varphi, \frac{\mathbf{r}+\mathbf{r}'}{2}\right)\frac{\partial^3}{\partial r_i\partial r_j\partial r_k}\delta(\mathbf{r}-\mathbf{r}') + \ldots \tag{32}$$

Here and below Latin indices varying from 1 to 3. Keeping only the first term of this expansion, one obtains from Eq. (32)

$$\frac{\partial^2\varphi}{\partial t^2} - c^2\Delta\varphi + m^2\varphi = \mathbf{f}(\varphi,\mathbf{r})\nabla\frac{\partial\varphi}{\partial t} + \frac{1}{2}\frac{\partial\varphi}{\partial t}\operatorname{div}\mathbf{f}(\varphi,\mathbf{r}). \tag{33}$$

In the simplest case, $\mathbf{f} = \nabla\varphi$ and

$$\frac{\partial^2\varphi}{\partial t^2} - c^2\Delta\varphi + m^2\varphi = \nabla\varphi\cdot\nabla\frac{\partial\varphi}{\partial t} + \frac{1}{2}\frac{\partial\varphi}{\partial t}\Delta\varphi. \tag{34}$$

Eq. (34) gives an example of scalar field with non-Lagrangian self-action.

Taking

$$E = \int_{-\infty}^{\infty}\left(u^3 - \frac{u_x^2}{2}\right)dx. \tag{35}$$

one gets

$$-\frac{\delta E}{\delta u} = 3u^2 + u_{xx} = \int_{-\infty}^{\infty} G(u,x,x',t)u_t(x',t)dx'. \tag{36}$$

For $G = \text{sgn}(x'-x)/2$ it leads to Korteweg - de Vries equation

$$u_t + 6uu_x + u_{xxx} = 0. \tag{37}$$

In the Coulomb gauge (div**A** = 0), the energy of the particles and electromagnetic field can be written as

$$E = \iiint \frac{\mathbf{E}^2 + \mathbf{H}^2}{8\pi} d^3\mathbf{r} + \sum_\alpha \frac{m_\alpha c^2}{\sqrt{1-\mathbf{v}_\alpha^2/c^2}} + \sum_{\alpha<\beta} \frac{Q_\alpha Q_\beta}{|\mathbf{r}_\alpha - \mathbf{r}_\beta|}, \tag{38}$$

where **E** and **H** are electric and magnetic fields, $m_\alpha$, $Q_\alpha$, $\mathbf{r}_\alpha$, and $\mathbf{v}_\alpha$ are the mass, charge, coordinate, and velocity of a particle with the number $\alpha$, and $c$ is the speed of light. Using Eq. (4), one obtains

$$L_1 = -\iiint \frac{\mathbf{E}^2 + \mathbf{H}^2}{8\pi} d^3\mathbf{r} - \sum_\alpha m_\alpha c^2 \sqrt{1-\mathbf{v}_\alpha^2/c^2} - \sum_{\alpha<\beta} \frac{Q_\alpha Q_\beta}{|\mathbf{r}_\alpha - \mathbf{r}_\beta|}. \tag{39}$$

Then Eq. (8) gives

$$\frac{1}{4\pi} F_{ai} = \int G_{abij}\left(\mathbf{F}, \frac{\partial \mathbf{F}}{\partial t}, \frac{\partial^2 \mathbf{F}}{\partial t^2}, \mathbf{r}_\alpha, \mathbf{v}_\alpha, \dot{\mathbf{v}}_\alpha, \mathbf{r}, \mathbf{r}', t\right) \frac{\partial F_{bj}(\mathbf{r}',t)}{\partial t} d^3\mathbf{r}' + \sum_\alpha G_{\alpha aij}\left(\mathbf{A}, \frac{\partial \mathbf{A}}{\partial t}, \frac{\partial^2 \mathbf{A}}{\partial t^2}, \mathbf{r}_\alpha, \mathbf{v}_\alpha, \dot{\mathbf{v}}_\alpha, \mathbf{r}, t\right) v_\alpha^j \tag{40}$$

and

$$\frac{d}{dt} \frac{m_\alpha v_\alpha^i}{\sqrt{1-\mathbf{v}_\alpha^2/c^2}} - Q_\alpha \sum_{\beta \neq \alpha} \frac{Q_\beta (r_\alpha^i - r_\beta^i)}{|\mathbf{r}_\alpha - \mathbf{r}_\beta|^3}$$
$$= -\int G_{\alpha aij}\left(\mathbf{F}, \frac{\partial \mathbf{F}}{\partial t}, \frac{\partial^2 \mathbf{F}}{\partial t^2}, \mathbf{r}_\alpha, \mathbf{v}_\alpha, \dot{\mathbf{v}}_\alpha, \mathbf{r}', t\right) \frac{\partial F_{aj}(\mathbf{r}',t)}{\partial t} d^3\mathbf{r}' + \sum_\beta G_{\alpha\beta ij}\left(\mathbf{F}, \frac{\partial \mathbf{F}}{\partial t}, \frac{\partial^2 \mathbf{F}}{\partial t^2}, \mathbf{r}_\alpha, \mathbf{v}_\alpha, \dot{\mathbf{v}}_\alpha, t\right) v_\beta^j \tag{41}$$

with the following antisymmetry conditions:

$$G_{ij}\left(\mathbf{A}, \frac{\partial \mathbf{A}}{\partial t}, \frac{\partial^2 \mathbf{A}}{\partial t^2}, \mathbf{r}_\alpha, \mathbf{v}_\alpha, \dot{\mathbf{v}}_\alpha, \mathbf{r}, \mathbf{r}', t\right) = -G_{ji}\left(\mathbf{A}, \frac{\partial \mathbf{A}}{\partial t}, \frac{\partial^2 \mathbf{A}}{\partial t^2}, \mathbf{r}_\alpha, \mathbf{v}_\alpha, \dot{\mathbf{v}}_\alpha, \mathbf{r}', \mathbf{r}, t\right), \tag{42}$$

$$G_{\alpha\beta ij}\left(\mathbf{A}, \frac{\partial \mathbf{A}}{\partial t}, \frac{\partial^2 \mathbf{A}}{\partial t^2}, \mathbf{r}_\alpha, \mathbf{v}_\alpha, \dot{\mathbf{v}}_\alpha, t\right) = -G_{\beta\alpha ji}\left(\mathbf{A}, \frac{\partial \mathbf{A}}{\partial t}, \frac{\partial^2 \mathbf{A}}{\partial t^2}, \mathbf{r}_\alpha, \mathbf{v}_\alpha, \dot{\mathbf{v}}_\alpha, t\right). \tag{43}$$

The simplest option of $G_{ij} = 0$, $G_{\alpha\beta ij} = Q_\alpha \delta_{\alpha\beta} \left[\partial A^j/\partial r^i (\mathbf{r}_\alpha) - \partial A^i/\partial r^j (\mathbf{r}_\alpha)\right]/c$ and

$$G_{\alpha ij} = \frac{Q_\alpha}{c} \left[\delta_{ij} \delta(\mathbf{r} - \mathbf{r}_\alpha) - \frac{\delta_{ij} |\mathbf{r} - \mathbf{r}_\alpha|^2 - (r^i - r_\alpha^i)(r^j - r_\alpha^j)}{4\pi |\mathbf{r} - \mathbf{r}_\alpha|^5}\right]$$

corresponds to common electrodynamics.

## 6. Conclusion

Thus equations of motion that conserve a given energy $E(\mathbf{q},\mathbf{v})$ were derived in this paper. They are Lagrange's equations with additional gyroscopic forces. In general case differential form of gyroscopic forces is not exact and such gyroscopic forces can not be described by additional terms in Lagrangian. Eq. (4) and Eq. (6) give an explicit expression of the Lagrangian through the energy. The generalized momenta and Hamiltonian can be found from the Lagrangian. The nonuse of variational principle is a remarkable feature of this approach. It allows obtaining more general equations of motion. The alternative axiomatics of the analytical mechanics and the field theory described in this paper is close to the axiomatics of thermodynamics, which is also based on conservation of energy.

## Acknowledgments

This work was supported by the World Class Institute (WCI) Program of the National Research Foundation (NRF) of Korea funded by the Ministry of Education, Science and Technology of Korea (NRF Grant Number: WCI 2011-001) and the Russian Scientific Foundation (project 14-50-00080).